\documentclass[letter]{aa}
\usepackage{graphicx}
\usepackage{txfonts}
\usepackage{hyperref}
\usepackage{graphicx}
\usepackage{bm}
\usepackage{amsmath}
\usepackage{amssymb}
\usepackage{amsfonts}
\usepackage{hyperref}
\usepackage{adjustbox}
\usepackage{natbib}
\usepackage{soul}
\usepackage{comment}

\bibpunct{(}{)}{;}{a}{}{,} 
\usepackage{bbold}
\hypersetup{colorlinks=true,citecolor=teal,linkcolor=teal,urlcolor=teal}
\usepackage{tabularx}
\usepackage{xspace}
\usepackage[normalem]{ulem}

\newcommand*{\Euclid}{\textit{Euclid}\xspace}

\definecolor{orange}{rgb}{1,0.5,0}
\definecolor{darkorange}{rgb}{0.69,0.33,0.13}
\definecolor{fidcol}{rgb}{0.7,0,0}     
\definecolor{mkcol}{rgb}{0.5,0,0.5}
\definecolor{mmcol}{rgb}{0.7,0.17,0.31}
\definecolor{dscol}{rgb}{0.6,0.1,0.2}
\definecolor{mccol}{rgb}{0.2,0.4,0.6}
\definecolor{darkgreen}{rgb}{0.05,0.5,0.06}
\definecolor{carnelian}{rgb}{0.7, 0.11, 0.11}
\definecolor{whatever}{rgb}{0.7,0.5,0.2}

\begin{document} 

    \title{Starlet higher order statistics for galaxy clustering and weak lensing}

   \author{Virginia Ajani
          \inst{1,2}
          \and Joachim Harnois-D{\'e}raps
          \inst{3}
          \and Valeria Pettorino
          \inst{1}
          \and Jean-Luc Starck
          \inst{1}
          }

   \institute{Université Paris-Saclay, Universit{\'e} Paris Cit{\'e}, CEA, CNRS, AIM, 91191, Gif-sur-Yvette, France
             \and 
             Institute for Particle Physics and Astrophysics, Department of Physics, ETH Zürich, Wolfgang Pauli Strasse 27, CH- 8093 Z\"urich, Switzerland
             \and 
             School of Mathematics, Statistics and Physics, Newcastle University, Herschel Building, NE1 7RU, Newcastle-upon-Tyne, UK
            \medbreak
            \email{vajani@phys.ethz.ch}}
             
\date{Received: 18 November 2022 ; Accepted: 15 March 2023}

\abstract{We present a first application to photometric galaxy clustering and weak lensing of wavelet-based multi-scale (beyond two points) summary statistics: starlet peak counts and starlet \emph{$\ell_1$-norm}. Peak counts are the local maxima in the map, and   $\ell_1$-norm is computed via the sum of the absolute values of the starlet (wavelet) decomposition coefficients of a map, providing a fast multi-scale calculation of the pixel distribution, encoding the information of all pixels in the map. We employ the {\it cosmo}-SLICS simulations sources and lens catalogues, and we compute wavelet-based non-Gaussian statistics in the context of combined probes and their potential when applied to the weak-lensing convergence maps and galaxy maps. We obtain forecasts on the matter density parameter $\Omega_{\rm m}$, the reduced Hubble constant $h$, the matter fluctuation amplitude $\sigma_8$,  and the dark energy equation of state parameter $w_0$.
In our setting for this first application, we consider the two probes to be independent. We find that the starlet peaks and the $\ell_1$-norm represent interesting summary statistics that can improve the constraints with respect to the power spectrum, even in the case of photometric galaxy clustering and when the two probes are combined.
  }
   
\maketitle

\section{Introduction}

Past, present, and future cosmological surveys, such as the Canada-France-Hawaii Telescope Lensing Survey (CFHTLenS) \citep{Heymans_2012}, the Kilo-Degree Survey (KiDS) \citep{heymans2020kids1000}, the Dark Energy Survey (DES) \citep{2019PhRvD..99l3505A, to2020dark}, Hyper SuprimeCam (HSC) \citep{2017AAS...22922602M}, \Euclid \citep{2011arXiv1110.3193L}, and the Rubin Observatory \citep{2009arXiv0912.0201L}, use weak gravitational lensing by large-scale structure and galaxy clustering as the main physical probes to explore the unknown properties of dark energy and dark matter. The correlation between the positions of foreground galaxies with the shapes of the source galaxies, also known as $3 \times 2$pt analysis, is very sensitive to the amplitude of matter clustering in the low-redshift universe and represents a powerful test of consistency between the growth of structure and the expansion history of the Universe \citep{PhysRevD.103.043503, PhysRevD.105.023520}. Since they bring complementary cosmological information, the combination of cosmic shear, galaxy clustering, and galaxy-galaxy lensing has proved to be very powerful to test cosmological models and modified gravity models and in enhancing the constraining power ( \citealt{PhysRevLett.99.141302, 10.1093/mnras/sty2168,heymans2020kids1000, 2020A&A...642A.191E, Tutusaus_2020} and references therein) and for dealing with systematic effects \citep{Mandelbaum:2012ay, 10.1093/mnras/stw2688,10.1093/mnras/sty2319, Joachimi_2021, DES:2021rex} as they are typically distinct and uncorrelated in different probes. Furthermore, the increasing precision reached with next-generation surveys will enable us to access the non-Gaussian part of cosmological signals, induced by the non-linear evolution of structure on small scales and low redshifts, which is not  
captured with second-order summary statistics alone. Specifically for weak lensing, a rich literature proposing several 
non-Gaussian statistics \citep{2023arXiv230112890E},  such as Minkowski functionals (e.g. \citealt{Kratochvil2012} and \citealt{Parroni2020}), higher order moments (e.g. \citealt{Petri2016} and \citealt{Gatti2020}), bispectrum (\citealt{Takada2004}, \citealt{Coulton_2019}), peak counts (\citealt{Kruse1999}, \citealt{die2010}, \citealt{PhysRevD.91.063507}, \citealt{Lin_2015}, \citealt{Peel_2017} \citealt{Martinet_2017}, \citealt{li2019}, \citealt{Ajani_2020} \citealt{zurcher2021dark}, \citealt{2022arXiv220406280A}),
Betti numbers \citep{Parroni_2021}, the scattering transform \citep{Cheng2020}, wavelet phase harmonic statistics \citep{Allys2020}, and machine learning-based methods (e.g. \citealp{Fluri2018} and \citealp{Shirasaki2021}), is catching the attention  of the  community.
The $\ell_1$-norm of wavelet coefficients of weak-lensing convergence maps has been proposed \citep{Ajani2021} as a new summary statistics for weak lensing  as it provides a unified framework to perform a multi-scale analysis that takes into account the information encoded in all pixels of the map.

Moreover, \cite{Minkowski_combined} have explored the addition of DES-like and LSST-like mock clustering maps to an analysis of  Minkowski functionals; they  find a significant improvement with respect to a weak lensing-only analysis. 
\cite{2022arXiv220309616K}   show that a  deep learning analysis of combined weak lensing and galaxy clustering at the map level can help break the degeneracy between cosmological and astrophysical parameters. As the wavelet-based statistics presented in \cite{Ajani_2020}, \citet{Ajani2021}, and \cite{Zurcher2022} were limited to weak lensing, we propose in this work a first application of such
statistics in the context of joint weak lensing and galaxy clustering
with the final goal of showing the potential of wavelet-based 
non-Gaussian statistics in the framework of probe combination.
The scope of this study is to extend the application of starlet peaks and $\ell_1$-norm to photometric galaxy clustering and its combination with weak lensing. 
Specifically, we employ lensing and clustering data from the \textit{cosmo}-SLICS simulations to build convergence and density maps from which we obtain various weak lensing and clustering predictions.
We perform a likelihood analysis using the power spectrum, the starlet peaks, and $\ell_1$-norm, and 
constrain the matter density parameter $\Omega_{\rm m}$, the reduced Hubble constant $h$, the matter fluctuation amplitude $\sigma_8$, and the dark energy equation of state parameter $w_0$. 

The letter is structured as follows. In Sect. \ref{sec: background} we provide some background definitions for weak lensing and photometric galaxy clustering. In Sect. \ref{sec:mock_data} we present the mock data, the catalogues, and our map-making procedure. We describe our analysis in Sect. \ref{sec:Analysis}.
In Sect. \ref{sec:results} we present and discuss the results, limitations, and perspectives of this work. We conclude in Sect. \ref{sec: Conclusions}.

\section{Background}\label{sec: background}

\subsection{Weak lensing}
The distortions caused by gravitational lensing in the original source image can be summarised in the distortion matrix: 

\begin{equation*}
\mathcal{A}=\left (
\begin{array}{cc}
1-\kappa -\gamma_1 &  -\gamma_2   \\
 -\gamma_2  & 1-\kappa +\gamma_1 \\
\end{array}
\right ) = (1 - \kappa)\left (
\begin{array}{cc}
1-g_1 &  -g_2   \\
 -g_2  & 1+g_1 \\
\end{array}
\right ).
\label{eq:Chap_2_Distortion_Matrix_Shear_Converrgence}
\end{equation*}

\noindent Here $\gamma=(\gamma_1, \gamma_2)$ is the shear, also defined in terms of the lensing potential $\psi$ as $\gamma_1 = \frac{1}{2}(\partial_1 \partial_1 - \partial_2 \partial_2) \psi$ and $\gamma_2 = \partial_1 \partial_2 \psi$, which describes the anisotropic distortion of the image; $\kappa$ is the convergence field, also defined as $\kappa =\frac{1}{2}\nabla^2 \psi$, which describes the isotropic dilation of the source; and $g$ is the reduced shear. The observed ellipticity $\epsilon^{\rm obs}$ is related with the intrinsic ellipticity of the galaxy $\epsilon^{s}$ and the reduced shear through the relation \citep{1997A&A...318..687S}

\begin{equation}
    \epsilon^{\rm obs} = \frac{\epsilon^s + g}{1 + g^{\star} \epsilon^s}
,\end{equation}

\noindent where $g^{\star}$ is the complex conjugate of $g$. In the  weak-lensing regime, where $|\gamma|, \kappa \ll 1$, this relation can be approximated as $\epsilon^{\rm obs} \approx \epsilon^s + \gamma$. Moreover, if the intrinsic ellipticity of galaxies has no preferred orientation, its expectation value 
vanishes, $\langle \epsilon^s \rangle=0$, and the observed ellipticity can be employed as an unbiased estimator of the reduced shear, as $\langle \epsilon^{\rm obs} \rangle \approx \gamma$ \citep{2015RPPh...78h6901K}. In this study, we work under these assumptions; including the effect of the intrinsic alignments goes beyond the purposes of this work.
The shear field can be related to the underlying matter density field using the relations connecting $\gamma$ and $\kappa$ to the lensing potential along with the fact that $\kappa$ is a line-of-sight integral of the matter density field. Working in Fourier space, we obtain an estimator of the convergence field starting from the shear field through the Kaiser-Squires inversion (KS) \citep{1993ApJ...404..441K} as

\begin{equation}
    \hat{\bf{\kappa}} = \frac{k_1^2 - k^2_2}{k^2}\hat{\gamma_1} + \frac{2k_1 k_2}{k^2}\hat{\gamma_2},
\end{equation}

\noindent where $k^2 = k_1^2 + k_2^2$.

\subsection{Photometric galaxy clustering}
As second-order statistics for photometric galaxy clustering, we employed the angular power spectra in harmonic space, $C^{\rm GG}(\ell)$, measured on the galaxy maps. 
Given the redshift distribution of the galaxies $n_{i}^{\rm G}(z)$ in each tomographic bin $i$ of a photometric survey, the average galaxy density in 
this redshift bin is given by \citep{2020A&A...642A.191E}

\begin{equation}
    \bar{n}_{i}^{\rm G}=\int_{z_{\rm min}}^{z_{\rm max}} {\rm d}z\,n_{i}^{\rm G}(z)\,.
\end{equation}

\noindent Then the radial weight function for a given tomographic bin $i$ for galaxy clustering can be defined as
\begin{equation}
    W_i^{\rm G}(z)=\frac{n_i^{\rm G}(z)}{\bar{n}_i^{\rm G}} \frac{H(z)}{c}\,.\label{eq:W_GCph}
\end{equation}

\noindent The observable spectrum is theoretically given by the  expression

\begin{equation}
    C_{ij}^{\rm GG}(\ell)=\int{\rm d}z\,\frac{W_i^{\rm G}(z)W_j^{\rm G}(z)}{H(z)r^2(z)}P_{\rm gg}^{\rm photo}\left[\frac{\ell+1/2}{r(z)},z\right]\,\label{eq:Cl_GCph}
,\end{equation}

\noindent where $P^{\rm photo}_{gg}(\ell,z)$ is the galaxy-galaxy power spectrum, which, under the assumption of a linear galaxy bias, is linked to the matter power spectrum $P_{\rm \delta \delta}(\ell, z)$ through the bias $b_{\rm g}$:

\begin{equation}
    P_{\rm gg}^{\rm photo}(\ell, z) = [b_{\rm g}^{\rm photo}(z)]^2P_{\rm \delta \delta}(\ell, z).\label{eq: P_gg_photo}
\end{equation}

\noindent In this first application we consider  only the auto-spectra ($i=j$) for our power spectrum measurements, and that the galaxy bias $b_{\rm g}$ is fixed in the simulations.

\section{Mock data}\label{sec:mock_data}
We modelled our non-Gaussian statistics with mock data constructed from the \textit{cosmo}-SLICS \citep{cosmo_SLICS_2019}, a suite of high-resolution N-body simulations consisting of 25 different cosmologies organised in a Latin hypercube obtained by varying the matter density parameter $\Omega_{\rm m}$ in the range $[0.10, 0.55]$, the reduced Hubble constant $h$ in the range $[0.60, 0.82]$, the combination between $\Omega_{\rm m}$ and the matter fluctuation amplitude $\sigma_8$, $S_8=\sigma_8\sqrt{\Omega_{\rm m}/0.3}$ in the range $[0.60, 0.90]$, and the dark energy equation of state parameter $w_0$ between $[-2.0, -0.5]$. Each gravity-only calculation evolves $1536^3$ particles in a
box of comoving side of 505 $h^{-1}$ Mpc with the Poisson solver {\sc cubep$^3$m} \citep{10.1093/mnras/stt1591}. These were then ray-traced under the Born approximation to construct the sources and lenses catalogues of $100$ ${\rm deg}^2$. To reduce the impact of cosmic variance, two simulations were provided for each cosmology with different initial conditions, each re-sampled to create 25 pseudo-independent light cones.

\subsection{Source catalogue}
For the weak lensing analysis we   used the KiDS-1000-like sources catalogue described
in \cite{Harnois_Deraps_2021}, which mimics the survey properties of the KiDS-1000 data described in \cite{10.1093/mnras/sty2271} and \cite{Hildebrandt_2021}. These catalogues are available for five different tomographic redshift bins, with a corresponding galaxy number density of $n_{\rm gal} = [0.62, 1.18, 1.85, 1.26, 1.31]$. For each source galaxy, the catalogue contains the positions  $(x, y)$ in arcmin and in redshift $z$, the values for the components of the true shear $(\gamma_1, \gamma_2)$ and the values for the components of the observed ellipticities $(\epsilon_1^{\rm obs}, \epsilon_2^{\rm obs})$. 

\subsubsection{Building convergence maps}
To get the convergence map, we used the KS inversion:  we projected the shear onto a $600^2$ Cartesian grid (resolution of 1 arcmin per pixel), which we then smoothed with a
Gaussian filter of width $0.7$ arcmin following \cite{10.1093/mnras/sty2271} to reduce the impact of potential empty pixels. This was implemented using the Python package \href{https://github.com/CosmoStat/lenspack}{\url{lenspack}}. We provide an example of our validation against the theoretical prediction in \autoref{AppendixC}. 

\begin{figure*}[h!]
    \centering
    \includegraphics[width=\textwidth]{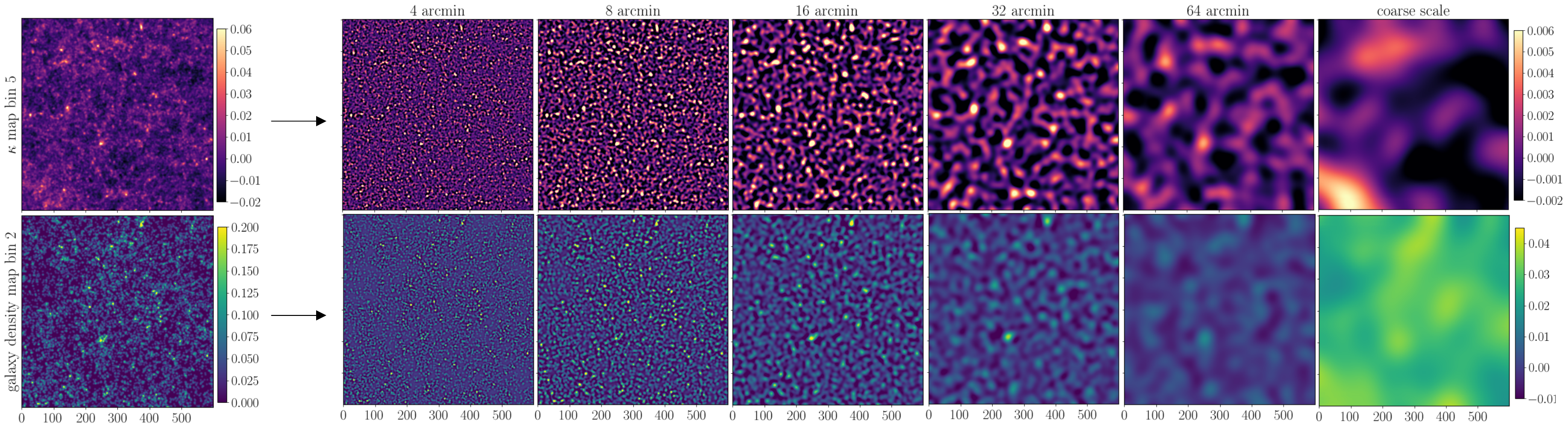}
    \caption{Starlet decomposition for a convergence map (top panel) and a galaxy map (bottom panel).    
    The value in arcmin corresponds to the resolution of the corresponding starlet scale. The resolution of the original map is 1 arcmin.
    }
    \label{fig:Chap_6_WL_GC_starlet_decomposition}
\end{figure*}

\subsection{Lens catalogues}\label{subsec:lense_catalogue}
To perform the galaxy clustering analysis, we 
employed the KiDS-1000-like lens catalogues of luminous red galaxies (LRGs) provided by the \textit{cosmo}-SLICS simulations. This simulated foreground LRG sample reproduces the characteristics described in \cite{Vakili_LRG_2020} and is split into four tomographic bins (see \citealt{burger2021revised}, for a recent study employing the same catalogues). Each catalogue contains the position and redshift of each galaxy. The LRGs trace the underlying projected mass sheets, hence their redshift distribution is heavily discretised in bins of $\Delta \chi$ = 257.5 Mpc/h. The catalogues are generated assuming linear galaxy bias, whose values and uncertainties are   from \cite{Vakili_LRG_2020} and summarised in \autoref{table_bias}.

\subsection{Building galaxy maps}

Starting from the lens catalogue described in Sect. \ref{subsec:lense_catalogue}, we built galaxy  maps for each light cone by assigning the galaxies onto a $600^2$ grid.
We validated our maps by comparing our measured power spectra against the theoretical prediction obtained with the Python library \href{https://pypi.org/project/pyccl/}{\url{pyccl}}. The shot noise of the galaxy maps was approximated as
Gaussian noise with a variance equal to the mean of each galaxy map.

\noindent As the map is intrinsically noisy (due to the presence of shot noise), while the theoretical prediction is obtained for a noiseless case, we computed the power spectrum of the noise map and then subtracted it from the power spectrum of the map. Since we did not have a model including non-linear bias, we started with a conservative analysis and included scales up to $\ell_{\rm max}=780$. An example of the comparison between our measurement on the simulated galaxy map and the corresponding theoretical prediction can be found in Fig. \ref{fig: GC_cl_validation}. 

\section{Analysis}\label{sec:Analysis}
\subsection{Summary statistics}\label{subsec: summary_statistics}

We compute our summary statistics on convergence maps for weak lensing and galaxy maps for galaxy clustering. When computing the power spectra, we smooth the noisy convergence and galaxy maps with a Gaussian kernel of size $\theta_{\ker}$ = 1 arcmin. 
For weak lensing we use 33 bins in the range $\ell = [180, 2040]$, which lies within the region where we are confident that the power spectra are in agreement with the theoretical predictions. For galaxy clustering, we use 14 bins in the range $\ell = [120, 780]$, following the \textit{pessimistic} setting of \cite{2020A&A...642A.191E}. These regions in $\ell$ are shown in \autoref{AppendixC}. For the non-Gaussian statistics, we filter the noisy maps with a starlet filter \citep{Starck2007}, a wavelet transform that allows for a fast multi-scale image analysis (see \autoref{AppendixA}). It has proved to be powerful for extracting cosmological information in the context of non-Gaussian statistics (e.g. \cite{Lin_2015}, \cite{Lin_2016},  \cite{Peel_2017}, \cite{Ajani_2020}, \cite{Ajani2021}). A visualisation of this decomposition is provided for both probes in Fig. \ref{fig:Chap_6_WL_GC_starlet_decomposition}. Following \cite{Ajani_2020}, we then computed peak counts on the starlet decomposed map, finding pixels with values larger than their eight neighbours. They can be quantified as a function of height, namely the pixel’s value, or the signal-to-noise ratio (S/N) \citep{Fan_2010, Maturi_peaks, shan-2017}. We compute the S/N as  
\begin{equation}\label{eq:singal_to_noise}
S/N \equiv \frac{(\mathcal{W} \ast \bm{m} )(\theta_\mathrm{ker})}{\sigma_{n}^\mathrm{filt} \bm{(\theta_\mathrm{ker})}},
\end{equation}

\noindent where $\mathcal{W}(\theta_\mathrm{ker})$ is the Starlet filter, $m$ is the noisy map, and  $*$ indicates the convolution operation. The quantity $\sigma_{n}^\mathrm{filt}$ is the standard deviation of the smoothed noise maps: for each weak-lensing map we estimate the shape noise as the standard deviation of the observed ellipticities from the source catalogue and for the galaxy clustering maps as the mean of each galaxy density maps. We estimate the noise level at each filter scale with automatic noise estimation from the multi-resolution support \citep{1998PASP..110..193S}.
On the same maps, we also compute the starlet $\ell_1$-norm, which is the absolute value of the sum of all wavelet coefficients for each scale of the decomposition within a chosen range of amplitude \citep{Ajani2021}. For both starlet peaks and $\ell_1$-norm we use for  inference four scales corresponding to [2\arcmin, 4\arcmin, 8\arcmin, 16\arcmin,  coarse], with 29 linearly spaced bins for each scale between the minimum and maximum values of each $S/N$ map.

\begin{figure}[!htbp]

\includegraphics[width=0.45\textwidth]{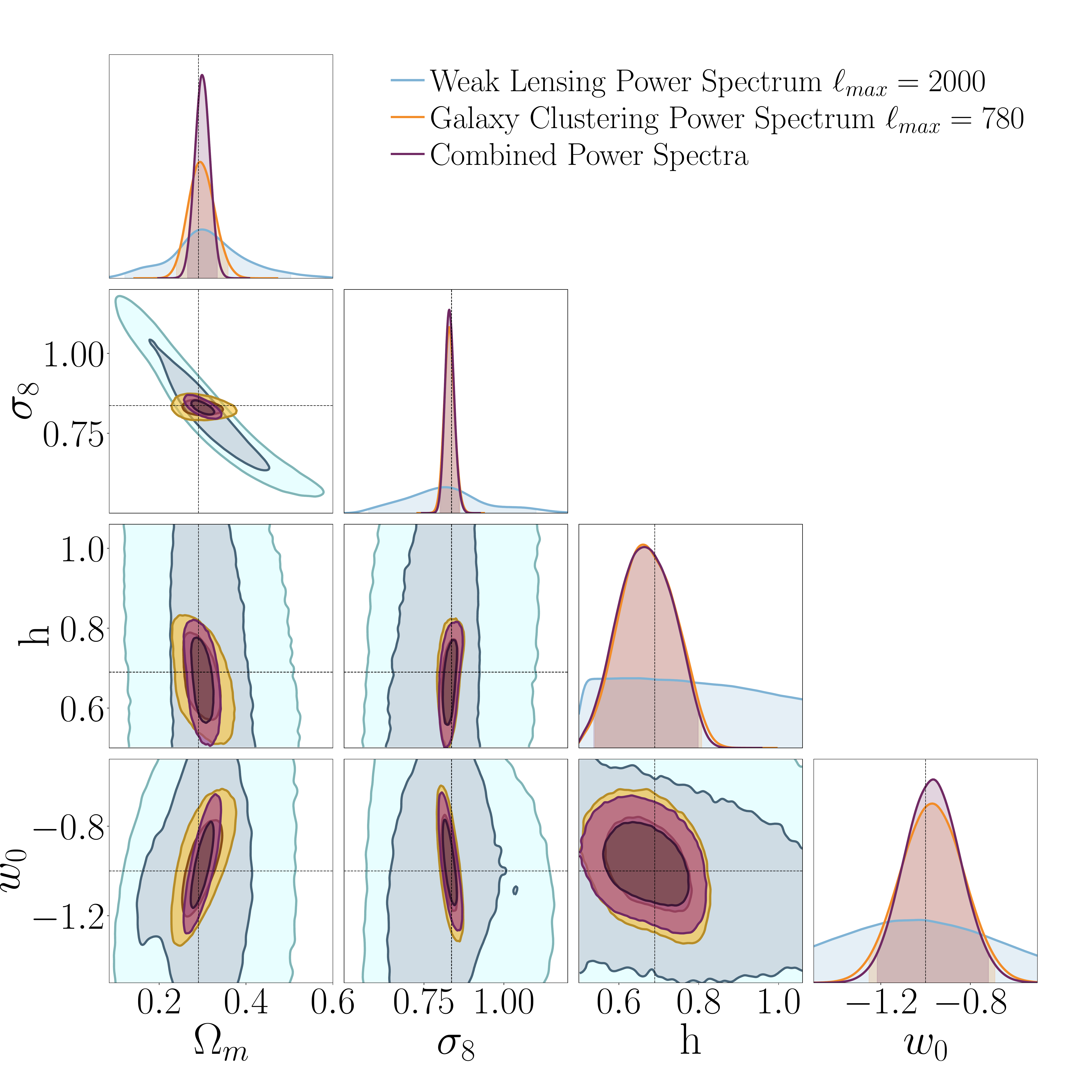}
\caption{Forecast contours at confidence levels of  68\% and 95\% using the weak-lensing power spectrum (light blue) compared to galaxy clustering power spectrum (orange) and their combination as independent probes (violet).
\label{fig: power_spectrum_GC_WL}}
\end{figure}

\subsection{Likelihood}
\label{sec:likelihood}
To perform Bayesian inference and obtain the probability distributions of the cosmological parameters we want to constrain, we use a Gaussian likelihood for a cosmology-independent covariance:

\begin{equation}\label{Likelihood_function}
\log \mathcal{L}(\mathbf{\theta})=\frac{1}{2} \sum_{\alpha} (d-\mu(\mathbf{\theta}))_{\alpha}^{T}C_{\alpha}^{-1}(d-\mu(\mathbf{\theta}))_{\alpha}.
\end{equation}

\noindent Here $d$ is the data array given by the average of the summary statistics over the different realisations at fiducial model, $C$ is the covariance matrix of the observable defined in the next section, $\mu$ is the model prediction as a function of the cosmological parameters $\theta$,  and $\alpha$ denotes the physical probe (i.e. weak lensing and galaxy clustering). At second order, the cross-correlation between the galaxy clustering and weak-lensing mock catalogues from two different surveys can be neglected if the overlap between the two surveys is sufficiently small \citep{Heymans_2021}. We are aware that in our case the galaxy density and convergence maps can in reality be correlated, as shown   in Fig. \ref{fig:Chap_6_WL_GC_starlet_decomposition}, where the structures of the galaxy map are traced by the lensing convergence map. 
For this first application, we   ignore these correlations that could actually bring additional information. For map level studies it has been shown that cross-probes can help in breaking degeneracies and in increasing the constraining power \citep{2022arXiv220309616K, Minkowski_combined}. 
We leave for a future study the inclusion of their  cross-correlation. The cosmological parameters are the ones that vary in the simulations, namely $\theta = [\Omega_{\rm m}, \sigma_8, h, w_0]$. To model the summary statistics at arbitrary cosmologies, we employ an interpolation with Gaussian processes regression \citep{10.5555/1162254}, using the \url{scikit-learn} Python package. For more details, the emulator used in this paper is the same as the one employed in \cite{li2019} and in \cite{2022arXiv220406280A}.\footnote{The code for the GP emulator is publicly available at \url{https://github.com/CosmoStat/shear-pipe-peaks}}

\subsection{Covariance matrix}

To compute the covariance matrix we employ the SLICS simulations \citep{10.1093/mnras/sty2319}. 
These are obtained as described in Sect. \ref{sec:mock_data} but are specifically designed for the estimation of covariance matrices: they consist of 800 fully independent $\Lambda$CDM runs in which the cosmological parameters are
fixed to the cosmology $[\Omega_{\rm m}, \sigma_8, h, n_s, \Omega_{\rm b}] = [0.2905, 0.826, 0.6898, 0.969, 0.0473]$  and the random seeds in the initial conditions are varied.
The covariance matrix elements are computed as
\begin{equation}\label{covariance_element}
C_{ij}=\sum\limits_{r=1}^{N} \frac{(x_{i}^{r} - \mu_i)(x_{j}^{r} - \mu_j)}{N-1},
\end{equation}

\noindent where $N=800$ is the number of observations, $x_i^{r}$ is the value of the $i$-{th} data element for a given realisation $r$, and $\mu_i$ is the mean over all realisations.
We take into account the loss of information due to the finite number of bins and realisations by adopting the estimator introduced by \cite{2007A&A...464..399H} for the inverse of the covariance matrix $C^{-1}=\frac{N-n_\mathrm{bins}-2}{N-1}C_{*}^{-1}$,
where $n_\mathrm{bins}$ is the number of bins and $C_{*}$ the covariance matrix computed for the power spectrum, the peak counts, and the $\ell_1$-norm, whose elements are given by Equation \eqref{covariance_element}. We area-rescale the covariance matrix to model the statistical accuracy of the  KiDS-1000 survey, with $A_{\rm KiDS-1000}=1000$ deg$^{2}$.

\label{Section:covariance}

\subsection{MCMC simulations and posterior distributions}
We explore and constrain the parameter space with the \url{emcee} package, the affine invariant ensemble sampler for Markov chain Monte Carlo (MCMC) introduced by \cite{2013PASP..125..306F}. We assume a flat prior for all parameters over the range modelled by the {\it cosmo}-SLICS, with the Gaussian likelihood defined in Equation (\ref{Likelihood_function}).  We use 120 walkers initialised in a tiny Gaussian ball of radius $10^{-3}$ around the fiducial cosmology $[\Omega_{\rm m}, \sigma_8, h, w_0]=[ 0.2905,  0.8364,  0.6898, -1]$.

\setlength{\tabcolsep}{26pt} 
\renewcommand{\arraystretch}{1.1}

\setlength{\tabcolsep}{10pt} 
\renewcommand{\arraystretch}{1.1}

\section{Results}\label{sec:results}

We first derive constraints with the power spectrum for both weak lensing and galaxy clustering, shown in Fig. \ref{fig: power_spectrum_GC_WL}. We see how, in our setting, the galaxy clustering alone outperforms the weak-lensing contours, breaking the degeneracies especially in the $(\Omega_{\rm m}, \sigma_8)$ plane. At the same time,   combining them brings additional information, improving the constraining power and the degeneracy break even more, underlying the complementarity of cosmological information encoded in the two probes.

\begin{figure}[!htbp]
    \includegraphics[width=\columnwidth]{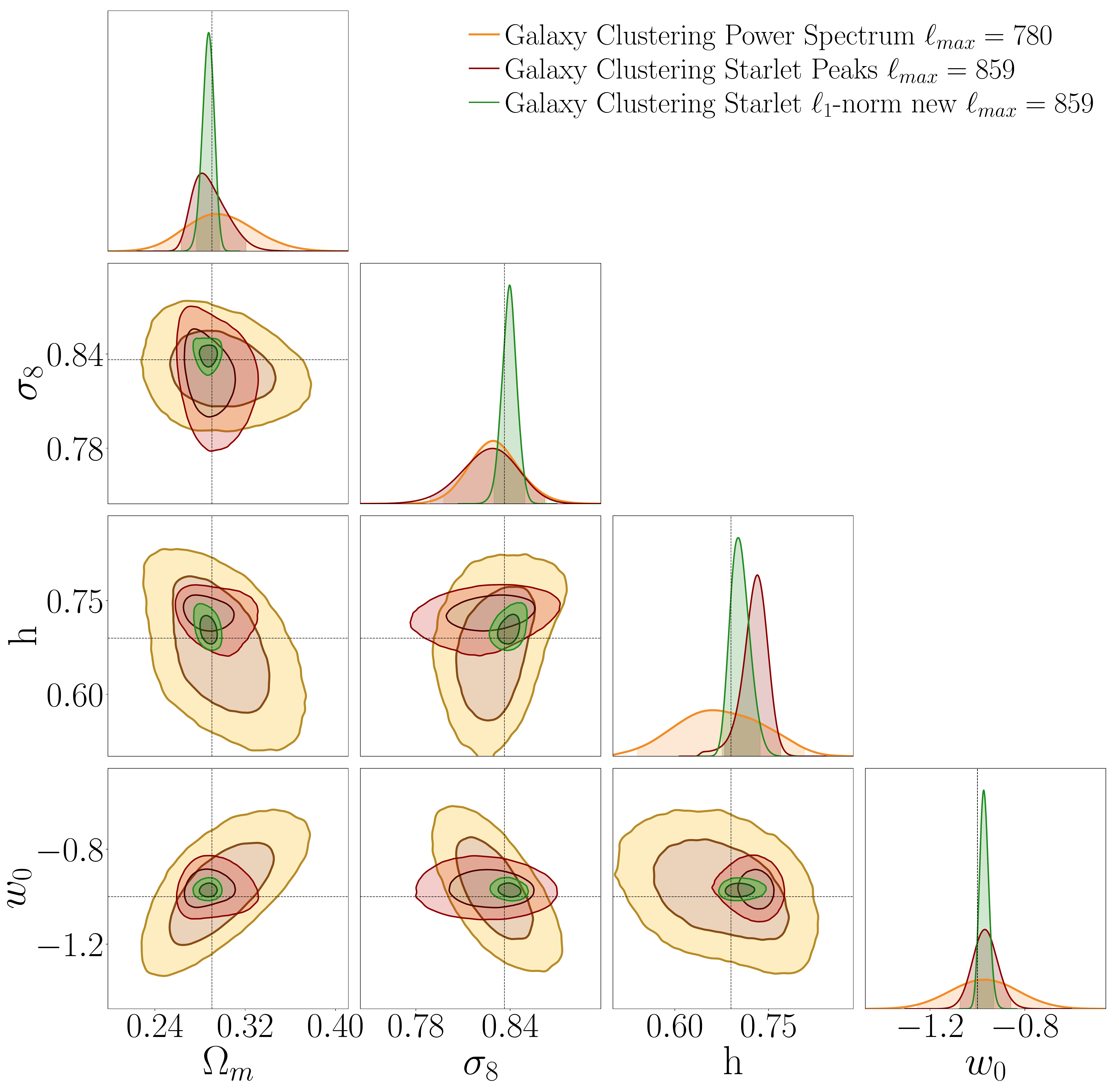} 
    \caption{Forecast contours at confidence levels of  68\% and 95\% using the power spectrum (orange), starlet peaks (light red), and the starlet $\ell_1$-norm(green) computed on galaxy density maps.}
    \label{fig:GC_l1_norm}
\end{figure}

\begin{figure}
    \centering
    \includegraphics[width=\columnwidth]{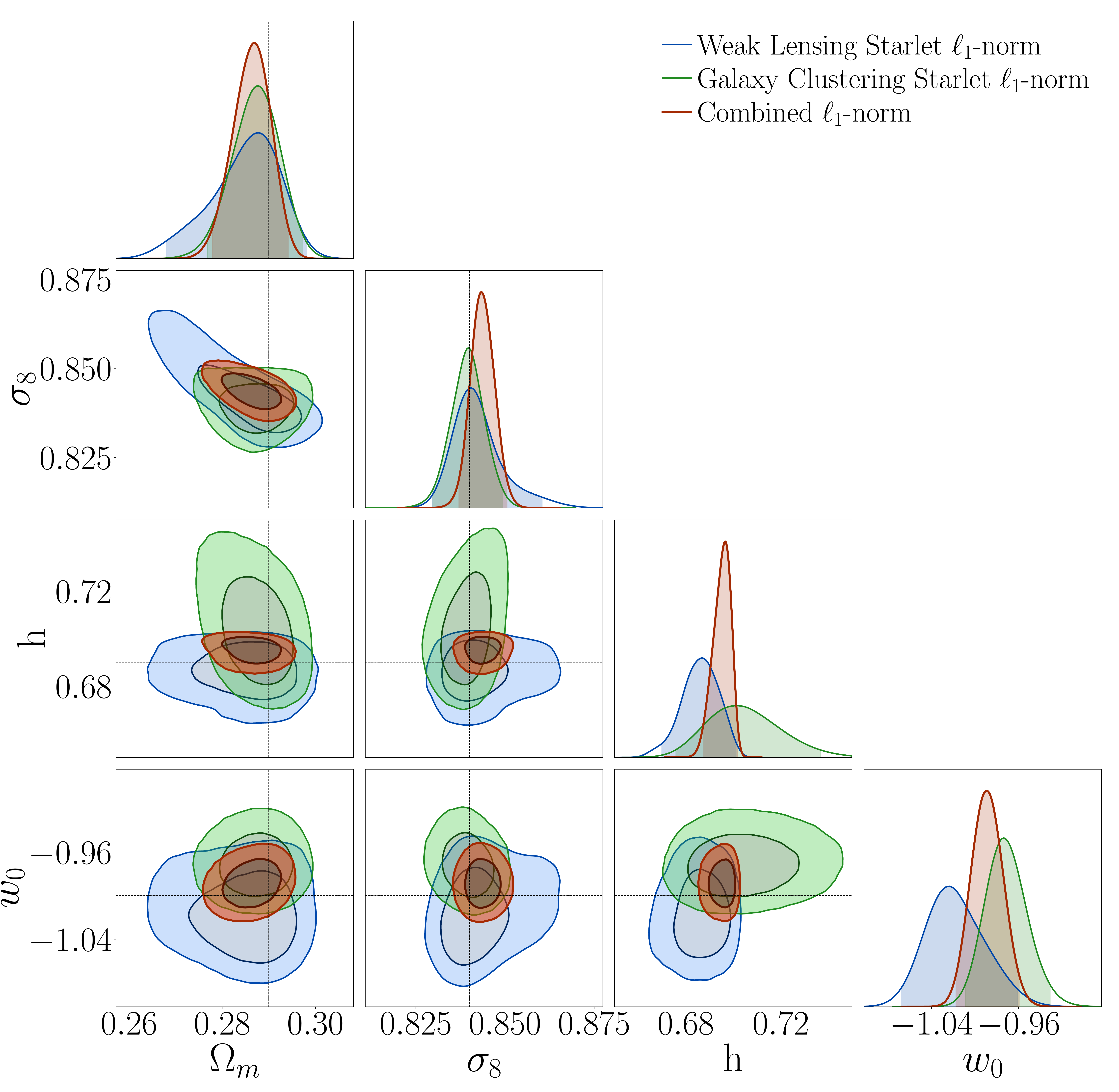}
    \caption{Forecast contours at confidence levels of  68\% and 95\% using the starlet $\ell_1$-norm measured on convergence maps (blue), on galaxy density maps (green), and their combination as independent probes (red).}
    \label{fig:l1_norm_combined}
\end{figure}

In Fig. \ref{fig:GC_l1_norm} we show the comparison among the constraints obtained using the galaxy clustering power spectrum, the starlet peak counts measured on galaxy density maps, and the galaxy clustering $\ell_1$-norm to explore this statistics in the context of galaxy clustering. We see how, as for weak lensing, the starlet peaks improve the constraining power with respect to the power spectrum and the $\ell_1$-norm outperforms the other two. We are aware that
the massive improvement in constraining power for galaxy clustering is very likely overestimated due to the fact that the galaxy bias in these simulations is kept fixed and assumed to be perfectly linear. In \autoref{AppendixB} we perform an analysis using the power spectrum 
with four different values of the galaxy bias parameters sampled from the measured uncertainty on these,  as reported in \autoref{table_bias}. The impact is major, where a $1\sigma$ shift in galaxy bias offsets the inferred $\sigma_8$ by $2\sigma$. To assess then the constraining power of the starlet peaks and the $\ell_1$-norm with respect to the power spectrum within our setting in a   fairer comparison, we show the results for galaxy clustering  alone, where the setting for all statistics in terms of maps employed and assumptions on the galaxy bias is the same. As in this setting the bias is fixed for all three statistics, we are more confident in claiming that   starlet peaks and the starlet $\ell_1$-norm both significantly improve the constraints, also in the case of galaxy clustering, with the $\ell_1$-norm outperforming the peaks. Given this result, to fully exploit the potential of multi-scale non-Gaussian statistics in the context of map-level probe combination, we get the constraints using the combination of starlet $\ell_1$
-norm for galaxy clustering and weak lensing. We show these results in Fig. \ref{fig:l1_norm_combined} for the weak-lensing starlet $\ell_1$-norm, the galaxy clustering starlet $\ell_1$-norm, and their combination. We see how, also in this context, the combination of the two probes might help in tightening the constraints, suggesting that this 
beyond-two-point statistics computed at the map level could be potentially very powerful for probe combination. We are aware that the relative importance of clustering will degrade once the bias is  allowed to vary. From the results of \autoref{AppendixB} we see how it is crucial to take this into account to avoid shifts on the inference contours for a real data application. We leave to a future study a deeper investigation of how we might include this uncertainty, for example by emulating the bias values across the uncertainties reported in \autoref{table_bias} and treating it as nuisance parameter.

\section{Conclusions}\label{sec: Conclusions}
In this Letter we proposed  using for the first time starlet-based 
non-Gaussian statistics on weak-lensing convergence maps and on galaxy maps to constrain cosmological parameters. We employed a mock source catalogue for weak lensing and a mock lens catalogue for galaxy clustering available from the {\it cosmo}-SLICS simulations provided with a galaxy bias estimated from the LRG set in the fourth data release of the Kilo-Degree Survey. We performed   a single probe and a combined 
non-Gaussian statistics analysis at the map level. We find that using the starlet-based statistics, such as starlet peak counts and the starlet $\ell_1$-norm also in the context of galaxy clustering, can potentially help in improving the constraints, and that this gain can be also obtained when combining weak lensing and photometric galaxy clustering. We are aware that there are many important aspects to explore before a real data application, such as the validity of linear bias approximation in the context of such higher order statistics. In \autoref{AppendixB} we show the results we obtained considering a different level of uncertainty on the bias and its impact on the constraints for the power spectrum.  
The next step was 
to build the framework to include galaxy-galaxy lensing analysis as well as cross-correlations between the two probes in the covariance matrix.
We expect these to help in breaking degeneracies and improving the constraining power. However, to exploit this step for a   real data application, several aspects need to be considered, such as observing conditions, masks, baryonic effects, and source clustering,  among others.  The goal of this first application represents a proof of concept to explore the applicability of these multi-scale map-based statistics already proven powerful for weak lensing also to galaxy clustering as a starting point for an analysis that accounts for the inclusion of systematic effects for both probes.

\begin{acknowledgements}
\noindent VA wishes to thank François Lanusse, for useful discussions and help during the first phase of development of the project. VA wishes to thank Tomasz Kacprzak and Alexandre Refregier for useful discussions. VA acknowledges support by the Centre National d’Etudes Spatiales and the project Initiative d’Excellence (IdEx) of Université de Paris (ANR-18-IDEX-0001) at the start and during the preparation of this work. VA acknowledges support by the grant \textit{Cosmological Weak Lensing and Neutral Hydrogen}, $200021\_192243$ of Prof. Alexandre Refregier. JHD acknowledges support from an STFC Ernest Ruther- ford Fellowship (project reference ST/S004858/1). The $N$-body simulations were enabled by the Digital Research Alliance of Canada (alliancecan.ca). 
\end{acknowledgements}

\bibliographystyle{aa} 
\bibliography{45510corr} 

\begin{appendix}

\section{Starlet transform}\label{AppendixA}

\begin{figure}[h!]
    \centering
    \includegraphics[width = \columnwidth]{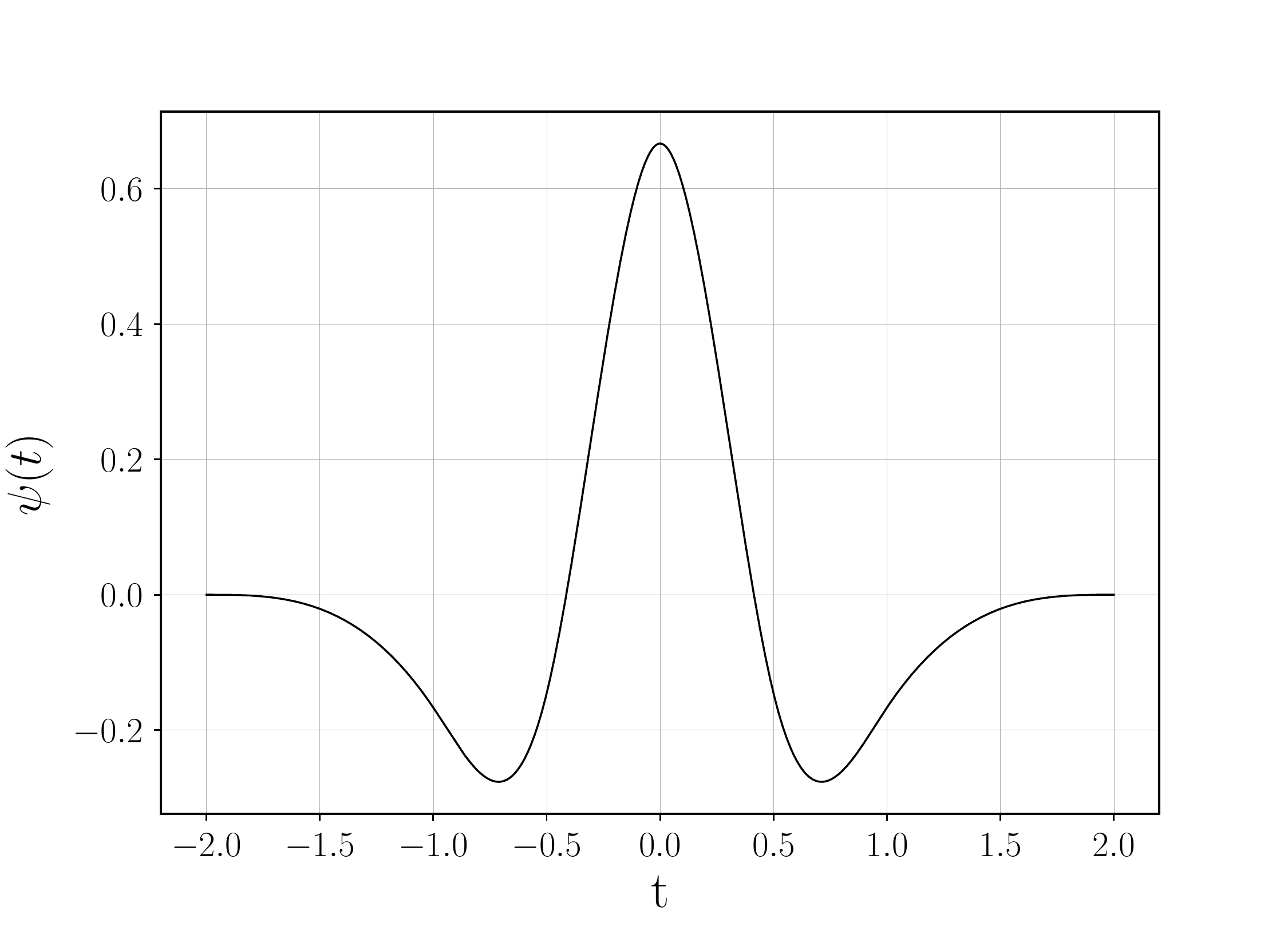}
    \caption{One-dimensional profile of the starlet function. }
    \label{fig:Starlet_1D_2D_profiles}
\end{figure}

We provide here a brief introduction to the starlet filter that we employed in this study to filter our input noisy maps. The starlet is a wavelet transform, namely a function that decomposes a signal into a family of scaled and translated functions. More specifically, wavelets are highly localised functions with vanishing mean since they 
satisfy the \textit{admissibility condition}  $\psi$  ($\int_{0}^{+ \infty} | \hat{\psi(k)} |^{2} \frac{\mathrm{d}k}{k} < + \infty$),  which implies that $\int \psi(x) \mathrm{d}x =0 $ (i.e. they integrate to zero over their domain). We can define the starlet $\psi$ in relation to its scaling function, which is a B-spline function $\phi$ of order 3,
\begin{equation}\label{Starlet_equation}
\psi(t_1,t_2)=4 \phi(2t_1,2t_2)-\phi(t_1,t_2)
,\end{equation}
with 
\begin{equation}
\phi (t) = \frac{1}{12}(|t-2|^3-4|t-1|^3+6|t|^3-4|t+1|^3+|t+2|^3)   
\end{equation}
and $\phi(t,t')=\phi(t)\phi(t')$ (for a complete description and derivation of the starlet transform algorithm, see  \citealt{Starck2007}). We show its 1D and 2D profiles in Fig. \ref{fig:Starlet_1D_2D_profiles}. When applying this transform in practice, an original $N \times N$ map \textit{I} is decomposed into a coarse version of   $c_{J}$ plus several images of the same size at different resolution scales \textit{j}:
\begin{equation}\label{eq:Starlet_decomposition}
I(x,y)=c_{J}(x,y) + \sum_{j=1}^{j_{\max}}w_j(x,y).
\end{equation}

\noindent Here the images $w_j$ represent the details of the original image at dyadic (powers of two) scales, corresponding to a spatial size of $2^j$ pixels and $J=j_{\max}+1$. Its shape emphasises round features, making it very efficient when dealing with peaks.

\section{Uncertainty on the bias}\label{AppendixB}
As the mock catalogues used in the galaxy clustering analysis are produced with a fixed galaxy bias, we provide here an estimate of the impact of the uncertainty of the galaxy bias on the parameter contours through MCMC forecasts. Specifically, we consider the best values and corresponding $\pm 1\sigma$ and $\pm 2\sigma$ found by \cite{Vakili_LRG_2020} for the sample of luminous red sequence galaxies for the fourth data release of the Kilo-Degree Survey. We provide in \autoref{table_bias} the values corresponding to Table 5 of \cite{Vakili_LRG_2020}. 

\begin{table}[h!]
    \centering
     \caption{Model constraints and uncertainties on the galaxy bias for the luminous red galaxies  from \citet{Vakili_LRG_2020} derived from the median and the $68\%$ confidence intervals of the marginalised posterior distributions.}
  \label{tab:my_label}
    \begin{tabular}{cccccc}
    bin&$-2\sigma$ & $-1\sigma$ & best & $+1\sigma$ & $+2\sigma$ \\ 
  \hline
   \hline
1 & 1.36&1.53   &1.70   &       1.88&   2.06\\ 
  \hline
2 & 1.46 & 1.59 & 1.72 & 1.86 & 2.00 \\ 
  \hline
3 & 1.62 & 1.68 & 1.74  &1.80   &1.86 \\ 
  \hline
4 & 1.85& 1.93  &       2.01&   2.09&   2.17 \\ 
\hline
\end{tabular}  \label{table_bias}
\end{table}

\begin{figure}[h!]
\centering
\includegraphics[width=0.9\columnwidth]{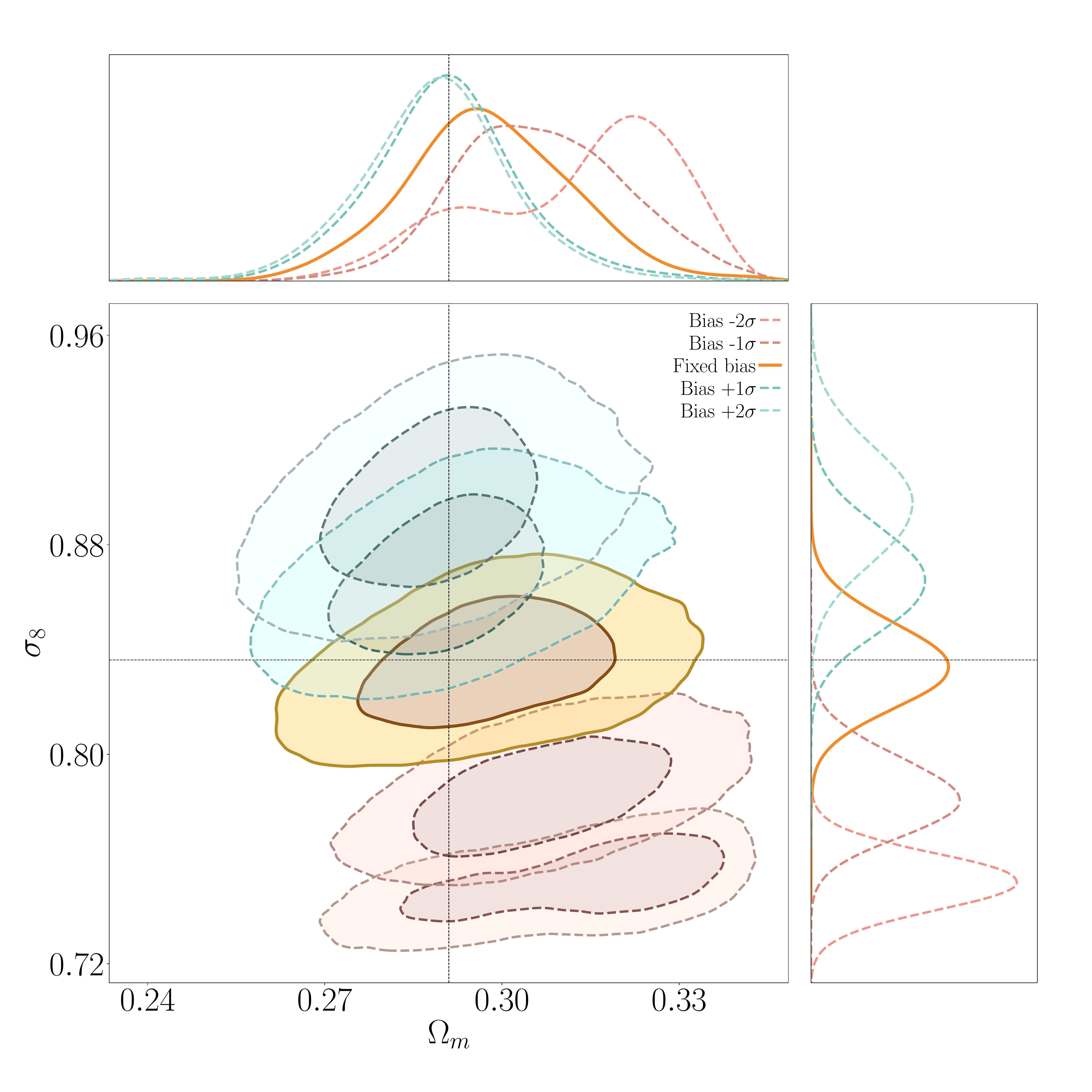}
\caption{Forecast contours at confidence levels of  68\% and 95\% using the power spectrum for four different uncertainties on the galaxy bias (dashed contours) compared to when the galaxy bias is fixed (continuous contours). The light blue dashed contours are   obtained using the columns `$+1\sigma$' and `$+2\sigma$' in \autoref{table_bias}; the light pink dashed contours using the columns `$-1\sigma$' and `$-2\sigma$'; and the orange continuous contours refer   to the column `best'. \label{fig:bias_mcmc}}
\end{figure}

\noindent In order to do so, we perform the inference analysis using the power spectrum computed for four different cases of uncertainty on the galaxy bias. We consider the bias \textit{unknown} using as mock data the power spectrum computed on simulations with respectively $+2\sigma$, $+1\sigma$, $-2\sigma$, and $-1\sigma$ uncertainty  on its value and for inference what was found as best fit (see `best' in \autoref{table_bias}) for the KiDS-1000 LRG set. We compare the results for these four settings with the case in which the galaxy bias is instead fixed (orange contours in Fig. \ref{fig:bias_mcmc}), namely the same in both mock data and simulations used for inference. We see how   the $\pm 1\sigma$ uncertainties in galaxy bias already induce a $2\sigma$ offset in the inferred $\sigma_8$, highlighting the importance of taking into account this uncertainty for real data application. In the future, we plan to perform  a deeper investigation of how the uncertainty on the bias impacts the results, for example by emulating and  marginalising over the effect.

\section{Map validation}\label{AppendixC}

\begin{figure}[h!]
\centering
\includegraphics[width=\columnwidth]{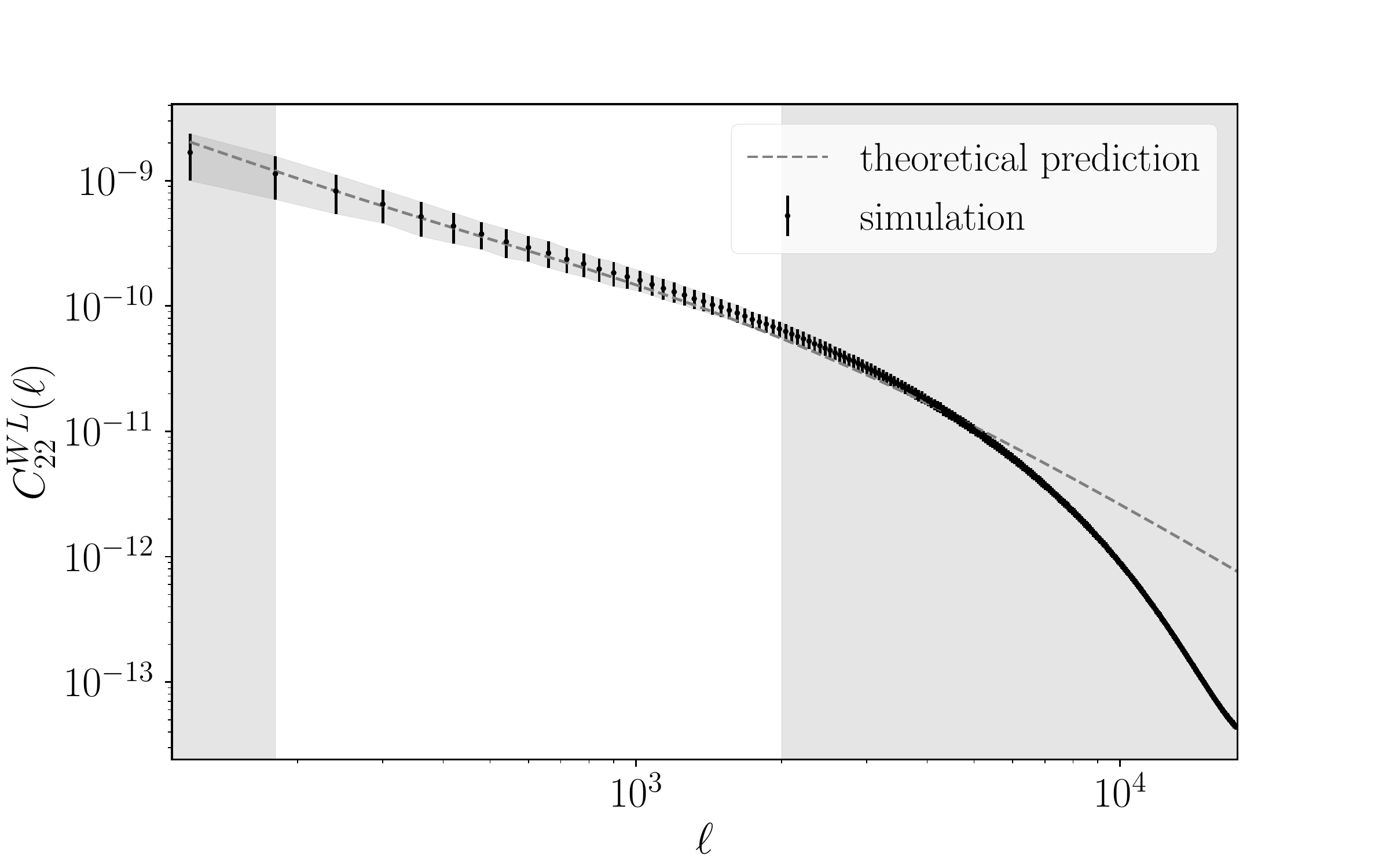}
\includegraphics[width=\columnwidth]{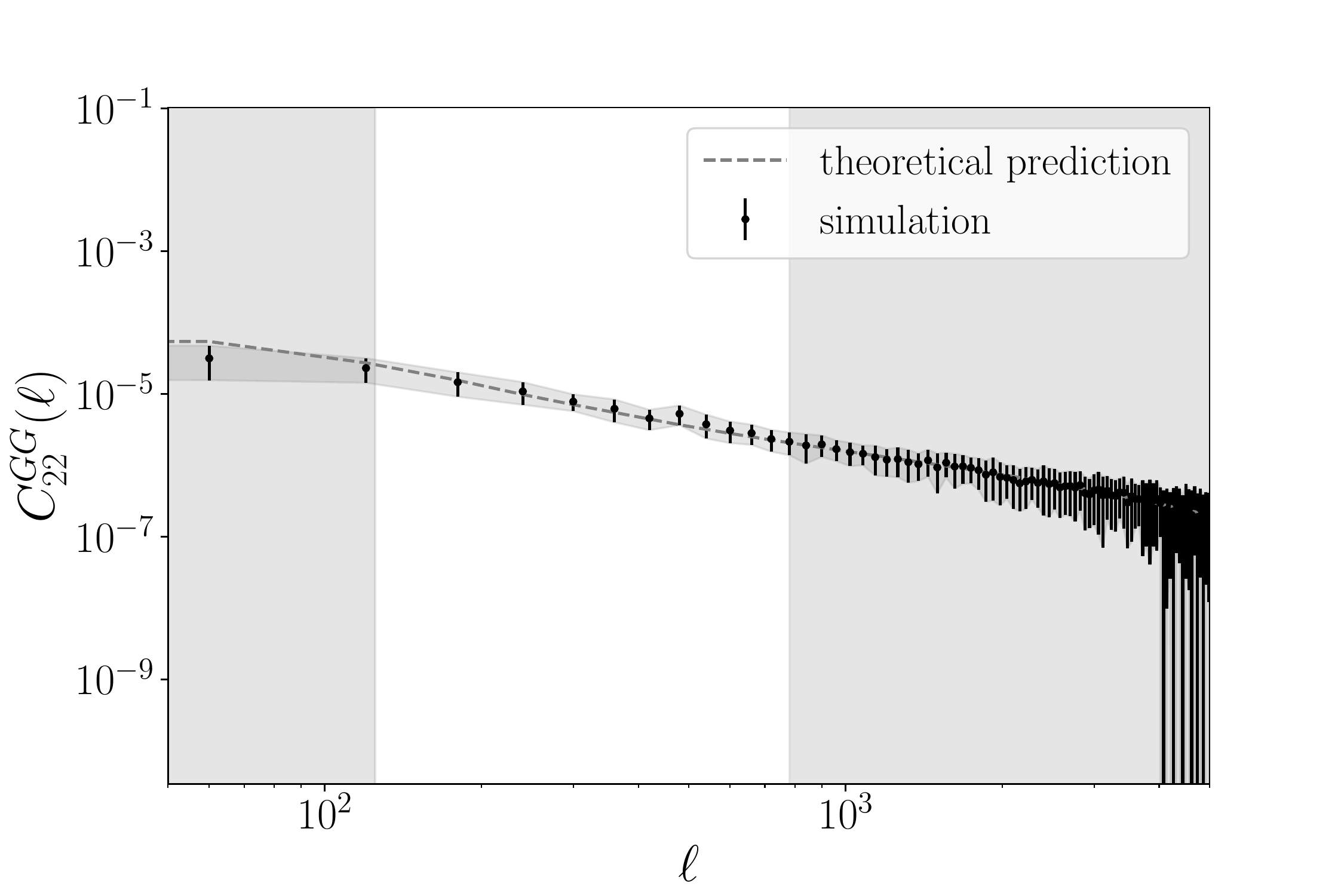}
\caption{Comparison between the power spectrum measured from maps built from the simulations and the corresponding theoretical prediction .\textbf{Top panel}: Weak-lensing power spectrum measured from convergence maps built from the simulations and the corresponding theoretical prediction obtained using \href{https://pypi.org/project/pyccl/}{\url{pyccl}}. An example of this comparison is shown for the fiducial cosmology of the {\it cosmo}-SLICS simulations of the third tomographic bin. \textbf{Bottom panel}: Same, but for the galaxy density map used for clustering.} \label{fig: GC_cl_validation}
\end{figure}

We validate the convergence and galaxy maps for the two probes that we built from the simulations by comparing our measured two-point statistics with the theoretical prediction. In order to do so we measure the power spectrum respectively on the convergence maps from the source catalogue and the galaxy density map from the lens catalogues built as described in Sect. \ref{sec:mock_data}. 
We show this in the top panel of Fig. \ref{fig: GC_cl_validation} for weak lensing and the bottom panel for galaxy clustering, where the black dots represent the angular power spectrum we measure from the galaxy map we built from the simulations averaged here across the 50 light cones, and the dashed line is the theoretical prediction using \href{https://pypi.org/project/pyccl/}{\url{pyccl}} \citep{Chisari_2019}. The grey areas indicate the scales excluded in the analysis. Concerning galaxy clustering, as the map is intrinsically noisy (due to the presence of
Poisson noise), while the theoretical prediction is obtained for a noiseless case, we compute the power spectrum of the noise and then subtract it from the power spectrum of the map. At high $\ell$ a noisy behaviour can be seen. The scales where this happens are excluded from the analysis, as we include multipoles up to $\ell_{\rm max}=780$.

\end{appendix}

\end{document}